\documentclass[aps,preprint,amsmath,amssymb,showpacs,amsfonts]{revtex4}
\usepackage{epsfig}
\usepackage{graphicx}
\usepackage{dcolumn}
\usepackage{bm}

\begin{document}

\title[New relations in  turbulence]
{New relations for correlation functions in Navier-Stokes
turbulence}

\author{Gregory Falkovich$^1$}
\author{Itzhak Fouxon$^2$}
\author{Yaron Oz$^2$}

\affiliation{$^1$ Physics of Complex Systems, Weizmann Institute
of Science, Rehovot 76100, Israel.\\$^{2}$  Raymond and Beverly
Sackler School of Physics and Astronomy, Tel-Aviv University,
Tel-Aviv 69978, Israel}

\begin{abstract}
We consider the steady-state statistics of turbulence  in the
inertial interval. The Kolmogorov flux relation (4/5-law) is shown
to be a particular case of the general relation on the
current-density correlation function. Using that, we derive an
analogous flux relation for compressible turbulence and a new
exact relation for incompressible turbulence.
\end{abstract}
\maketitle
\section{Introduction}
The Kolmogorov flux relation  is a rare exact analytic result one
has for the correlation functions of the velocity $\bm v$ in the
Navier-Stokes equation. It is traditionally derived considering
quadratic invariants like kinetic energy in an incompressible flow
(or squared vorticity in two dimensions). Conservation of the
kinetic energy, $\int v^2/2$, by an unforced Euler equation means
that one can define the energy flux in $k$-space and write the
continuity equation for the energy spectral density
\cite[]{Frisch}:
\begin{eqnarray}
&& \Pi_k=-{1\over 8\pi^2}\int d^3r{\sin({\bf k\cdot r})\over
r}{\partial\over\partial r_i}\left[{r_i\over
r^2}{\partial\over\partial r_j}\langle u_ju^2\rangle\right]\
.\nonumber\\&&{\partial \langle |v_k|^2\rangle\over2\partial
t}+{\rm div}\,\Pi_k=0,\ \bm u(\bm r)\equiv \bm v({\bm r})-\bm v(0).
\label{flux0}
\end{eqnarray}
Stationarity under the action of a large-scale force and
small-scale viscosity means  the constancy of the energy flux
$\Pi_k$ over intermediate wavenumbers
\cite[]{Kolmogorov,Frisch,Gawedzki}:
\begin{equation}\nabla_i\langle
u_{i}u^2\rangle=-4\nabla_i\langle v_{i}(\bm r)v_{j}(\bm
r)v_{j}(0)\rangle=-4\epsilon\ .\label{flux1}\end{equation} That
can also be written for the third moment of the longitudinal
velocity difference, $u_l=({\bf u}\cdot{\bf r}/r)$:
\begin{equation}
\langle u_l^3\rangle=-12\epsilon r/d(d+2)\
.\label{flux10}\end{equation} Here $d$ is the space
dimensionality. For $d=3$ one obtains $\langle
u_l^3\rangle=-4\epsilon r/5$, which is why the flux relation is
sometimes called 4/5-law. Analogous relations are derived for the
passive scalar turbulence, magnetized and helical flows
\cite[]{Yaglom,Chandrasekhar,PouquetPolitano,Pod1,Pod2,Galtier,Hel1,Hel2,Hel3}.
If, however, the respective conservation law is non-quadratic (as
the energy in compressible turbulence, for instance) then it is
believed that the analogs of (\ref{flux1}) are absent. In
addition, there are no analogs of this relation for velocity
moments of the orders $n$ different from three. Experiments
demonstrate that $\langle u_l^n\rangle\propto r^{\zeta_n}$ with
the scaling exponents $\zeta_n$, which are larger than one but
smaller than $n/3$, see e.g. \cite{Frisch,FalkovichSreenivasan}.

In this letter, we re-interpret the Kolmogorov relation in terms
of currents and densities of the conserved quantities, which
allows us to derive an analog for a compressible case and a new
fundamental relation for an incompressible case.
\section{General relation}
Consider a general class of classical field dynamics,
\begin{eqnarray}&&
\partial_t q^a+\nabla\cdot \bm j^a=f^a,
\end{eqnarray}
where $q^a, a=1,...,N$ are densities, $\bm j^a$ are currents and
$f^a$ are the external random source fields. These equations
describe local conservation laws and  provide a canonical form for
the effective hydrodynamic description of a slow evolution of
large-scale perturbations. Since the zero wave-number component of
the density is conserved, low wave-number components evolve slowly
by continuity, see e.g. \cite{Forster}. The equations are closed
via a constitutive relation that expresses currents in terms of
the densities as a series in gradients:
\begin{eqnarray}&&
j_i^a=F^a_i(\{\rho\})+\sum_{jb}G^a_{i, jb}(\{\rho\})\nabla_j
\rho^b+\ldots, \label{exp}
\end{eqnarray}
where dots stand for higher order terms involving more
derivatives. The zero-order, reactive, term leads to a
conservative dynamics while the first-order term describes
dissipation. For fluid mechanics, consideration of higher order
terms in Eq.~(\ref{exp}) is usually unnecessary so we limit
ourselves to the following general class, which also contains the
Navier-Stokes equation:
\begin{eqnarray}&&
\partial_t q^a+\frac{\partial F_i^a}{\partial r_i}=f^a-\frac{\partial }{\partial r_i}
\left(\sum_{jb}G^a_{i, jb}(\{\rho\})\nabla_j \rho^b \right).
\label{dyn}
\end{eqnarray}
We  assume the standard mathematical formulation of the problem of
turbulence where the forcing term $f^a$ is random and its
statistics is stationary, spatially homogeneous and isotropic \cite{Frisch}.
This implies
 that the same
properties hold for the steady state statistics of $q^a$.  The
correlation length of the force will be denoted below by $L$.

The derivation of the Kolmogorov type relation for Eq.~(\ref{dyn})
proceeds as follows. Consider the steady-state condition $
\partial_t\langle q^a(0, t)q^a(\bm r, t)\rangle=0$ (no summation over $a$).
Employing the dynamical equation (\ref{dyn}) and using statistical
symmetries, one finds
\begin{eqnarray}&&
0=\partial_t  \langle q^a(0, t)q^a(\bm r,
t)\rangle=-2\frac{\partial}{\partial r_i} \langle q^a(0, t)
F_i^a(\bm r, t)\rangle \nonumber\\&& +2\langle q^a(0, t) f^a(\bm
r, t)\rangle\nonumber
-\frac{\partial }{\partial r_i}\left\langle q^a(0, t)
\left(\sum_{jb}G^a_{i, jb}( \{\rho\}(\bm r, t))\nabla_j \rho^b(\bm
r, t)\right)\right\rangle. \nonumber
\end{eqnarray}
We consider the limit of large correlation length $L$ of the
forcing. That allows one to consider $r$ much smaller than $L$ yet
still large enough, so that the last term becomes negligible as
containing higher order of spatial derivatives. Because $r\ll L$
we have $f^a(\bm r, t)\approx f^a(0, t)$ and $\langle q^a(0, t)
f^a(\bm r, t)\rangle\approx \langle q^a(0, t) f^a(0,
t)\rangle\equiv \epsilon_a$, where the constant $\epsilon_a$ is
the mean input rate of $ (q^a)^2$.  Hence we obtain for the
single-time correlation function
\begin{eqnarray}&&
\nabla_i\langle q^a(0) F_i^a(\bm r)\rangle=\epsilon_a.
\end{eqnarray}
Assuming in addition isotropy one finds
\begin{eqnarray}&&
\langle q^a(0) F_i^a(\bm r)\rangle=\frac{\epsilon_a r_i}{d}\ .
\label{general}
\end{eqnarray}

\section{Particular cases}

A simple example is the passive scalar turbulence when some
substance with the density $q$ and diffusivity $\kappa$ is carried
by a flow with the velocity  $\bm v$. The current is $\bm j=q\bm
v-\kappa\nabla q$. The flux (Yaglom) relation is as follows:
$\langle q(0) q(\bm r)\bm v(\bm r)\rangle={\epsilon \bm r}/{d}$.

Another example is the turbulence of a barotropic fluid where the
pressure  $p(\rho)$  is a function of the fluid density $\rho$
only. In this case, ${\bf q}=(\rho{\bf v},\rho)$, $F^i_j=\rho
v_iv_j+p(\rho)\delta_{ij}$ and $\bm j^4=\rho\bm v$. The equations
have the form
\begin{eqnarray}&&
\partial_t\rho+\nabla\cdot(\rho\bm v)=0,\ \ \
\partial_t(\rho v_i)+\partial_j (\rho v_iv_j+p\delta_{ij}) \label{a1}\\&&
=-\partial_j \left[G^i_{j, kb}(\{\rho\})\nabla_k
\rho^b\right]+f^i, \label{a2}
\end{eqnarray}
where we took into account that $\bm j^4=\rho \bm v$ is exact  to
all orders. The source often has the form $f^i=\rho \nabla_i \Phi$
that corresponds to an external potential $\Phi$ (the analysis
below can be easily generalized to other types of forcing as
well). The exact form of $G^i_{j, kb}$ is not important below but
its presence is necessary for a steady state. Indeed, the
 rhs of Eq.~(\ref{a2}) breaks the energy
conservation; the steady state holds due to the balance of the
forcing that pumps fluctuations into the system and the
dissipation that erases them. Now we  use the general relation
(\ref{general}) to derive the new relation for the compressible
turbulence described by Eqs.~(\ref{a1})-(\ref{a2}). The
application of the relation to $q^4=\rho$ with $\epsilon_4=0$
gives $\langle \rho(0, t)\rho(\bm r)v_i(\bm r)\rangle=0$. In fact
this result holds for any relation between $r$ and $L$. Indeed,
the steady-state condition $\partial_t \langle \rho(0, t)\rho(\bm
r, t)\rangle=0=\partial_i\langle \rho(0)\rho(\bm r)v_i(\bm
r)\rangle$ applied to the general (isotropic) form $\langle
\rho(0)\rho(\bm r) v_i(\bm r)\rangle=A(r)r_i$,  gives $A=C
r^{-d}$, so that regularity at the origin requires $C=0$.

Application of Eq.~(\ref{general}) to $q^j$ with $j=1, 2, 3$ gives
a non-trivial relation
\begin{eqnarray}&&
\sum_j\left\langle \rho(0)v_j(0)\left[\rho(\bm r)v_j(\bm r)v_i(\bm
r)+p(\bm r)\delta_{ij} \right]\right\rangle=\frac{\epsilon r_i}{d}
\label{Kolmcompr}
\end{eqnarray}
where $\epsilon$ is defined in this case as $\langle \rho(0)\bm
v(0)\cdot \bm f(0)\rangle=\epsilon$  (we summed over $j$ to get a
more symmetric result). To the best of our knowledge, the formula
(\ref{Kolmcompr}) is new and presents a desired analog of
Kolmogorov relation for the compressible turbulence see e. g.
\cite{Kritsuk}. Probably, this relation was not derived before
because it demands considering the steady-state condition for the
fourth-order correlation function $\langle \rho(0, t)v_j(0,
t)\rho(\bm r, t)v_j(\bm r, t)\rangle$, while usually in trying to
find Kolmogorov type relations one considers steady state
conditions for the second moment, like in magnetohydrodynamics by
\cite{PouquetPolitano}. Note in passing that one can choose $q$ as
the energy density and obtain yet another relation analogous to
(\ref{TP0}) below.

In the incompressible limit, $\rho=const$ and $\nabla\cdot\bm
v=0$, the pressure term is zero; again, since it is a
divergence-free vector that must be regular at the origin
\cite[see][]{LL}. In this case, (\ref{Kolmcompr}) is reduced to
(\ref{flux1}),
\begin{eqnarray} &&
\langle v_{i}(\bm r)v_{j}(\bm r)v_{j}(0)\rangle=\epsilon
r_i/d\,,\label{d1}
\end{eqnarray}
and Eq.~(\ref{flux10}) is implied. Hence  (\ref{Kolmcompr}) is
indeed a general form of the Kolmogorov relation for an arbitrary
Mach number. As we see, from a general viewpoint, the relation
follows from the stationarity of the pair correlation function of
the momentum density rather than from the energy spectral density.
Indeed, $\epsilon$ in (\ref{Kolmcompr}) is the input rate of the
squared momentum   and not that of the energy, which coincide (up
to the factor $1/2$) only in the incompressible case.

We have shown how the Kolmogorov relation exploits the momentum
conservation. Now in the same way we shall exploit the energy
conservation and derive a new fundamental relation for
incompressible turbulence. Energy conservation in the
incompressible case means that the Navier-Stokes equation can be
written as a continuity equation for the kinetic energy:
\begin{eqnarray}&&
\frac{\partial}{\partial t}\frac{v^2}{2}=
-\frac{\partial}{\partial
r_i}\left[v_i\left(\frac{v^2}{2}+p\right)\right]+ \bm f\cdot\bm
v+\nu v_i\nabla^2 v_i. \label{a9}
\end{eqnarray}
That means that one can choose $q=v^2$ which turns the general
form (\ref{general}) into a fifth-order relation. The only
difference from the third-order relation (\ref{d1}) is that the
pressure term is now nonzero. Note first that the condition
$\partial \langle v^4\rangle/\partial t=0$ gives the single-point
pressure-velocity correlation function:
\begin{equation}\langle v^2v_i\nabla_ip\rangle=\langle
\bm f \cdot\bm v v^2\rangle+\nu \left\langle v^2 v_{i} \nabla^2
v_{i}\right \rangle\ .\label{singlet}\end{equation} For the
different-point moment, Eq. (\ref{general}) gives
\begin{eqnarray}&&\!\!\!\!
\frac{\partial}{\partial r_{i}}\left\langle v_{i}(\bm
r)\left[\frac{v^2(\bm r)}{2}+p(\bm r)\right]v^2(0)\right\rangle
=\langle \bm f(\bm r)\cdot\bm v(\bm r)v^2(0)\rangle+\nu
\left\langle v^2(0) v_{i}(\bm r) \nabla^2 v_{i}(\bm r)\right
\rangle\,.\label{TP0}
\end{eqnarray}
This relation can be further simplified by decoupling small-scale
and large-scale fields. Force and velocity are large-scale fields
that change over the scale $L$ so that one can put $\langle \bm
f(\bm r)\cdot\bm v(\bm r)v^2(0)\rangle=\langle \bm f\cdot\bm v
v^2\rangle$ at $r\ll L$ as we did in deriving (\ref{general}).
Velocity differences and derivatives are small-scale fields that
change respectively over the scale $r$ and the viscous length
$\eta$ see e. g. \cite{Frisch}. In particular, the local energy
dissipation, $\nu v_i(\bm r) \nabla^2 v_i(\bm r)$, is a
small-scale field with the correlation radius $\eta$. When the
distance $r$ is in the inertial interval, $r\gg \eta$, one can
decouple large-scale and small-scale fields: $\nu\left\langle
v^2(0)v_i(\bm r) \nabla^2 v_i(\bm r)\right\rangle\approx
-\epsilon\langle v^2\rangle$. We also denote $\bm V=\bm v(\bm
r)+\bm v(0)$ and present
\begin{eqnarray}&&
4\left\langle v_{i}(\bm r) v^2(\bm
r)v^2(0)\right\rangle=-\left\langle u_{i}\left[v^2(\bm
r)-v^2(0)\right]^2\right\rangle=-\left\langle u_{i}(\bm u\cdot \bm
V)^2\right\rangle\nonumber\\&&\approx -\left\langle
u_{i}u^2\right\rangle\left\langle
V^2\right\rangle\left\langle\cos^2\alpha\right\rangle= -2\langle
v^2\rangle\left\langle u_{i}u^2\right\rangle=8\langle
v^2\rangle\epsilon\bm r/d\ .\label{dec}
\end{eqnarray}
Here $\alpha$ is the angle between $\bm u$ and $\bm V$ and in the
second line we assumed that the vectors $\bm u$ and $\bm V$ are
weakly correlated and can be decoupled when the distance $r$ is in
the inertial interval, $r\ll L$. In the last equality we used the
Kolmogorov relation (\ref{flux1}). We thus find that in the
inertial range, $\eta\ll r\ll L$, Eq.~(\ref{TP0}) is reduced to
\begin{eqnarray}&&
\frac{\partial}{\partial r_{i}}\left\langle v_{i}(\bm r)p(\bm
r)v^2(0)\right\rangle =\langle \bm f\cdot\bm v
v^2\rangle-2\epsilon\langle v^2\rangle\equiv D\,.\label{TP1}
\end{eqnarray}
We thus obtain the fundamental relation on the pressure-velocity
correlation function which is the counterpart of the Kolmogorov
relation (\ref{flux10}):
\begin{eqnarray}
\left\langle \bm v(\bm r)p(\bm r)v^2(0)\right\rangle = D\bm
r/d\,.\label{TP2}
\end{eqnarray}
That relation can be tested experimentally and numerically.  From
a formal viewpoint, (\ref{Kolmcompr},\ref{TP1}) are particular
cases of the Hopf equations which express the stationarity of the
correlation functions. Hopf equations are generally not closed;
they impose some relations between different structure functions
but do not allow to find them see e. g. \cite{vy}. On the
contrary, our relations (\ref{Kolmcompr},\ref{TP1}) allow one to
find the correlation functions that are fluxes. Taking divergence
of (\ref{TP2}) gives $\langle v^2(0)v_{i}(\bm r)\nabla_ip(\bm
r)\rangle=D$ in the inertial interval, compare it with the
single-point expression (\ref{singlet}).

Inverting the incompressibility condition, $\Delta p=-\nabla_i
v_j\nabla_jv_i$, one expresses pressure:
\begin{eqnarray}&&
p(\bm r)=\frac{1}{(d-2)\sigma_d} \int\frac{d\bm r'}{|\bm r-\bm
r'|^{d-2}} \frac{\partial^2 u_i(\bm r')u_j(\bm r')}{\partial
r'_i\partial r'_j},\ \ \sigma_d=\frac{2\pi^{d/2}}
{\Gamma(d/2)}\label{pres}\,,
\end{eqnarray}
where $u_i(\bm r')=v_i(\bm r')-v_i(0)$. That allows one to present
(\ref{TP2}) as an integral relation on the fifth-order three-point
moment of the velocity field (we denote ${\bm\hat r}' =\bm r'
/r'$):
\begin{eqnarray}&&
\frac{1}{\sigma_d}\int\frac {d\bm r'}{r'^d} \left\langle v^2(\bm
r)\bm v(0)\left[u^2(\bm r')-d\left[\bm u(\bm r')\cdot\bm\hat
r'\right]^2\right]\right\rangle=D\bm r /d.
\label{TP01}\end{eqnarray} We expressed it via the differences in
the square brackets to make it explicit that the integrand is
regular at the origin.

Note that the force term in (\ref{TP1}) is the input rate of the
squared energy which for a general force statistics cannot be
expressed via the energy input rate. In the particular case of a
white Gaussian force, such expression is possible: $\langle
f_iv_iv^2\rangle=(1+2/d)\epsilon\langle v^2\rangle$ and
$D=(2/d-1)\epsilon\langle v^2\rangle$. Remark that the direct
energy cascade, which is studied here, is absent for $d=2$. For 2d
inverse energy cascade, one  derives in a similar way:
$\left\langle \bm v(\bm r)p(\bm r)v^2(0)\right\rangle =
\epsilon_4\bm r/2$ where $\epsilon_4$ is the dissipation rate of
the squared energy due to a large-scale sink. For the direct
cascade of vorticity $\omega=\nabla\times\bm v$ in 2d, analogous
flux relations on the correlation functions $\left\langle \bm
v(\bm r)\omega^n(\bm r)\omega^n(0)\right\rangle$ do not contain
pressure and were derived and analyzed by \cite{FL}.

Let us discuss the validity limits of (\ref{dec}) where we
neglected cumulants like the structure function $\langle\bm
uu^4\rangle_c$. Such structure function scales as $r^{\zeta_5}$
and is subleading at sufficiently small $r$ since $\zeta_5>1$. In
other words, there exists the scale $\ell$, below which the
decoupling is possible. We ask now how $\ell$ and $\zeta_5$ may
depend on the only parameter that enters Navier-Stokes equation,
the space dimensionality $d$. For $d=3$ we expect $\ell\simeq L$.
Usually, in statistical physics \cite[]{stan} and passive scalar
theory \cite[]{Kra74,CFKL,kaz,FGV} the statistics is getting
closer to Gaussian and decoupling improves when $d$ grows.
Somewhat counter-intuitively, one may expect a different behaviour
in turbulence because of diminishing role of pressure.  To
understand how the pressure depends on $d$, consider how the
identity $\Delta p=-\nabla_i v_j\nabla_jv_i$
 behaves when  $d$ increases while the velocity
components are kept fixed. The Laplacian has $d$ terms of
different signs. Assuming that in the limit $d\to\infty$ those
terms can be considered independent, their sum grows like
$\sqrt{d}$. The rhs contains $d^2$ terms and grows like $d$ so
that the pressure is expected to grow as $p\propto\sqrt{d}$. This
is slower than $v^2\propto d$ so that assuming $d$ large one may
neglect the pressure contribution into the correlation functions.
Consider, for instance, the tensor of the fourth moment of the
velocity difference:
\begin{eqnarray}&&
\langle u_iu_ju_ku_l\rangle=A(r){\hat r}_i{\hat r}_j{\hat
r}_k{\hat
r}_l+C(r)\left[\delta_{ij}\delta_{kl}+\delta_{ik}\delta_{jl}+
\delta_{il}\delta_{jk}\right]\nonumber\\&& + B(r)
\left[\delta_{ij}{\hat r}_k{\hat r}_l+\delta_{ik}{\hat r}_j{\hat
r}_l+\delta_{il}{\hat r}_k{\hat r}_j+\delta_{jl}{\hat r}_k{\hat
r}_i+\delta_{jk}{\hat r}_i{\hat r}_l+\delta_{kl}{\hat r}_i{\hat
r}_j\right]\ .\nonumber
\end{eqnarray}
At $d\to\infty$ the difference between longitudinal and
transversal velocity components is expected to decrease, as
follows from the analysis of the second moment where
incompressibility and isotropy requires $\langle
u_\perp^2-u_l^2\rangle= r\langle u_l^2\rangle'/(d-1)$. Comparing
$\langle u_l^4\rangle=A+6B+3C$ with $\langle u_\perp^4\rangle=3C$,
we see that  $A,B\ll C$ at $d\to\infty$. Now, we generalize for
$d$ dimensions the relation between the velocity and pressure
correlation functions, derived by \cite{Hill} for three
dimensions, and remarkably find out that $C$-terms cancel out:
\begin{eqnarray} && \langle
p^2\rangle=\frac{d^2-1}{12} \int_0^{\infty} A(r')\frac{dr'}{r'}-
\frac{d-1}{3}\int_0^{\infty} \frac{dr'}{r'}
\left[A(r')+3B(r')\right].
\end{eqnarray}
That means that indeed $\langle p^2\rangle$ grows slower than
$\langle u^4\rangle$ as $d\to\infty$.
Remind that generally the correlations functions containing $p$
and $u$ may scale differently yet considering the limit
$d\to\infty$ {\it at fixed $r$} one may neglect the pressure term
in (\ref{TP0}). That would give \begin{equation}\nabla_i
\left\langle v_{i}(\bm r) v^2(\bm r)v^2(0)\right\rangle=2\langle
\bm f\cdot\bm v v^2\rangle-2\epsilon\langle v^2\rangle=
4\epsilon\langle v^2\rangle/d \label{New}\end{equation} (for
Gaussian pumping) which is much different from $\epsilon\langle
v^2\rangle$ given by (\ref{dec}). That means that at $d\to\infty$
there may exist an interval of scales, $L\gg r\gg\ell$, where
cumulants are comparable to the reducible part so that $\ell$ is a
crossover scale between (\ref{TP1}) and (\ref{New}).   That would
mean that $\ell$ decreases with $d$ while $\zeta_5\to1$ as
$d\to\infty$. Moreover, at $d\to\infty$, the same analysis can be
done for all odd moments: $\zeta_n=1$ for $n\geq1$. We thus come
to the suggestion that the scaling of incompressible turbulence in
the limit $d\to\infty$ may be the same as the scaling of Burgers
turbulence. This would not be that surprising since a single
incompressibility condition imposed on $d$ velocity components
 is expected to be less restrictive  as $d$ grows, so that
flow configurations close to shocks give the main contribution to
the moments. Note that for shock-like configurations, one cannot
decouple $\bf u$ and $\bm V$ as done in deriving (\ref{TP1}).
Technically, neglecting pressure, makes all quantities $\int
v^{2n}\,d{\bf r}$ to be integrals of motion of the Euler equation
so that the linear scaling of all odd velocity moments express the
constancy of fluxes of these integrals of motion, like in Burgers,
see e.g. \cite{CFG}. Holder inequality then requires the linear
scaling for even moments too, which corresponds to an extreme
non-Gaussianity of the small-scale velocity statistics.
Physically, pressure is a non-local field which couples different
regions in space and is expected to act like ``an intermittency
killer'' as remarked by \cite{Kra},  see also \cite{GN,Nel}. One
way to find the exponents at finite $d$ may be a large-$d$
expansion, which must thus be very different from that used by
\cite{stan,CFKL,kaz}: one needs to start here from a Burgers-like
limit rather than from a Gaussian statistics. That explains, in
particular,  why no substantial simplification was found  and why
pressure terms were not small in the large-$d$ perturbation theory
that starts from a Gaussian statistics \cite[]{FFR}.

Needless to say that the tendency of the exponents to approach
unity with $d$ growing remains purely hypothetical on that level
of analysis. However, if true, it means that the degree of
non-Gaussianity of the statistics of a single velocity component
grows with $d$, which agrees with the numerical comparison between
three and four dimensions made in the remarkable work by Gotoh and
co-authors \cite{Gotoh}. Scalar quantities made out of vector
products contain the sum of $d$ terms; one cannot generally
conclude whether their statistics gets closer to Gaussian as $d$
increases because of the competition between increasing
non-Gaussianity of a single term and the averaging over $d$ terms.

To conclude, the formulas (\ref{Kolmcompr},\ref{TP1},\ref{TP2}) is
the main result of this work.

\begin{acknowledgments}
The work  was supported  by the Israeli Science Foundation, the
Deutsch-Israelische Projektkooperation, the US-Israel Binational
Science Foundation, the Minerva Foundation  and  the
German-Israeli Foundation. G.F. thanks M. Nelkin, U. Frisch, T.
Gotoh and G. Boffetta for useful remarks.
\end{acknowledgments}

\bibliographystyle{jfm}

\end{document}